\documentstyle[12pt]{article}
\begin{document}
\begin{center}
{\bf{INFRARED REGION OF QCD AND CONFINING STRINGS}}

\vspace{1.0cm}

R.Parthasarathy{\footnote{e-mail
address:sarathy@imsc.ernet.in}} \\
The Institute of Mathematical Sciences \\
C.I.T. Campus, Taramani Post \\
Chennai 600 113, India. \\
\end{center}

\vspace{1.0cm}

{\noindent{\it{Abstract}}}

\vspace{0.5cm}

Gauge field configurations appropriate for the infrared region
of QCD are proposed in a submanifold of $su(3)$. Some
properties of the submanifold are presented.  
Using the
usual action for QCD in the absence of quarks, confinement of
these configurations is realized as in the London theory of
Meissner effect. Choosing a representation for the monopole
field strength, a string action corresponding to the effective
gauge theory action in the infrared region, is obtained. This
confining string action contains the Nambu-Goto term, extrinsic
curvature action and the Euler characteristic of the string
world sheet.

\vspace{0.5cm}


\newpage  

{\noindent\bf{I.Introduction}}

\vspace{0.5cm}

QCD, the gauge theory of strong interaction based on the gauge
group $SU_c(3)$, exhibits asymptotic freedom [1] in the high
energy region and in this region, the success of the
perturbation theory (due to the small coupling strength of the
QCD running coupling constant) has been supported by
experimental results [2]. In the infrared region where the
coupling becomes strong, the perturbation theory cannot be
used and  
the mechanism of confinement of the
gluons and the quarks is not completely understood. t'Hooft
[3] and Mandelstam [4] suggested that magnetic monopoles must
play a crucial role in the mechanism of confinement. Color
confinement is possibly due to a dual Meissner effect [4,5,6].
This suggests that in the infrared region, some other
variables other than $A^a_{\mu}$ may be more relevant, and
Mandelstam [7] suggested that the monopole plasma is probably
a way of parameterizing the ground state of the confining
phase. The idea of 'tHooft [3] is to make an Abelian
projection and this has been discussed by Mandelstam [7]. In
particular, Kondo [8] has examined this for $SU(2)$ gauge
theory to obtain an Abelian projected effective field theory.
Here the non-Abelian degrees of freedom are integrated and the
relation between the non-Abelian gauge fields and the monopole
configuration is not clear. Recently, Faddeev and Niemi [9]
have proposed a set of variables for describing the infrared
region of 4-dimensional $SU(2)$ and $SU(N)$ gauge theory,
continuing the earlier works of Corrigan et.al., [10] and Cho
[11]. They [9] propose a non-linear sigma model action as
relevant in the infrared region of Yang-Mills theory. The
author has recently proposed [12] gauge field configurations
for the infrared region of QCD with $SU(3)$ as the gauge
group, in which the magnetic confinement of gluons and quarks
have been realized as in the London theory of Meissner effect.
A dual version produces the electric confinement as well,
thereby explicitly demonstrating the scenario proposed by
Nambu [6].

\vspace{0.5cm}

While this field theoretic approach to confinement is based
upon the attempts to identify relevant gauge field
configurations as in [9, 12], there is another approach based
upon the string representation as pointed out by Polyakov
[13]. Starting from the action for compact $U(1)$ gauge
theory, he [13] obtains the action for the confining strings
as the rigid string action (which contains the extrinsic
curvature action besides the Nambu-Goto action) in the large
Wilson loop limit. It is to be noted here that compact $U(1)$
gauge theory {\it{must include}} the monopole configurations [14]. 
This
important connection [13] between the gauge fields (including
monopole configurations) and strings has been extended to
4-dimensional case by Diamantini, Quevedo and Trugenberger
[15] by introducing the antisymmetric Kalb-Ramond fields [16]
interacting with the monopoles. These studies [13,15] reveal
two important features. First, the string action so obtained
contains besides the Nambu-Goto action, the extrinsic
curvature action, earlier proposed by Polyakov [17] and
independently by Kleinert [18] as relevant to describe QCD
strings. Second, the coefficient of the extrinsic curvature
action is found to be negative, as desired by Kleinert and
Chervyakov [19] to improve the stability of the theory.

\vspace{0.5cm}

It is the purpose of this paper to first identify the relevant
gauge field configurations in the infrared region of QCD and
to obtain an action [12] which {\it{naturally}} contains the
monopole configurations,  
exhibiting confinement and {\it{then}} to
obtain a string representation of {\it{this}} low energy
effective action by going to the Euclidean version. 
In Section.II, the proposed [12] $SU(3)$
gauge field configuration appropriate in the infrared region
of QCD are briefly reviewed and some properties of the
submanifold are obtained in Section.III. An  
effective action exhibiting confinement is
obtained in Section.IV. The string representation of this
action (in Euclidean space) is obtained in Section.V. The
results are summarized in Section.VI.  

\vspace{0.5cm}

{\noindent{\bf{II. Gauge Field Configurations for the Infrared
Region of QCD.}}}

\vspace{0.5cm}

The suggestions of 'tHooft [3], Mandelstam [4,7] and Polyakov
[13] that the low energy region of QCD must contain monopole
configurations to provide a mechanism of confinement, can be
implemented by proposing the $SU(3)$ gauge field
configurations $A^a_{\mu}$ satisfying 
\begin{eqnarray}
D^{ab}_{\mu}{\omega}^b \ =\ {\partial}_{\mu}{\omega}^a + g\
f^{acb}\ A^c_{\mu}\ {\omega}^b &=& 0,
\end{eqnarray}
where ${\omega}^a$ is an $SU(3)$ octet vector $\in\ SU(3)/Z_3$
and  
are chosen such that
\begin{eqnarray}
{\omega}^a\ {\omega}^a &=& 1, \nonumber \\
d^{abc}\ {\omega}^b\ {\omega}^c &=& \frac{1}{\sqrt{3}}\
{\omega}^a,
\end{eqnarray}
with $d^{abc}$ as the symmetric Gell-Mann $SU(3)$ tensors.
Now, in the infrared region, the gauge group manifold is not
the complete $su(3)$ group manifold spanned by arbitrary
${\omega}^a$'s but a submanifold spanned by those
${\omega}^a$'s satisfying (2) and the relevant gauge field
configurations $A^a_{\mu}$ are {\it{determined}} by (1). The
second relation in (2) is very special to $SU(3)$ and will be
crucial in what follows.   

\vspace{0.5cm}

A solution to (1) is
\begin{eqnarray}
A^a_{\mu} &=& C_{\mu}\ {\omega}^a\ -\ \frac{4}{3g}\ f^{abc}\
{\omega}^b\ {\partial}_{\mu}{\omega}^c,
\end{eqnarray}
where $C_{\mu}$ is arbitrary. To show that (3) solves (1), we
made use of the following relations
\begin{eqnarray}
{\omega}^a\ {\partial}_{\mu}{\omega}^a &=& 0, \nonumber \\
d^{abc}\ ({\partial}_{\mu}{\omega}^b)\ {\omega}^c &=&
\frac{1}{2\sqrt{3}}\ {\partial}_{\mu}{\omega}^a,
\end{eqnarray}
which follow from (2), and the relation [21]
\begin{eqnarray}
f^{abc}f^{edc}&=&\frac{2}{3}({\delta}_{ae}{\delta}_{bd}-
{\delta}_{ad}{\delta}_{be}) + d^{aec}d^{dbc} - d^{adc}
d^{bec},
\end{eqnarray}
for the $f$ tensors. We find a useful relation for the
${\omega}^a$'s in (2) as
\begin{eqnarray}
\frac{4}{3}\ f^{abc}f^{edc}\ {\omega}^b {\omega}^e\
{\partial}_{\mu}{\omega}^d &=& -{\partial}_{\mu} {\omega}^a,
\end{eqnarray}
which will be used subsequently. The gauge field configuration
(3) and the submanifold determined by (2) are proposed as
relevant to describe the infrared region of QCD.    

\vspace{0.5cm}

It is therefore very essential to show, in view of [3,4,7,13],
that the proposed configurations (3) admits monopoles
necessary for confinement. This will be shown in the 
Section IV. Before this, we note two important consequences that
follow from (1). A mass term $m^2\ A^a_{\mu}A^a_{\mu}$ for the
gauge fields usually cannot be added to the lagrangian due ot
its gauge non-invariance. Since under a gauge transformation
the field changes as $\delta A^a_{\mu}\ =\
D^{ab}_{\mu}{\omega}^b$, if we consider gauge transformations
within the submanifold (2), then such a term is allowed in
view of (1). Second, if we choose a Lorentz covariant gauge
${\partial}_{\mu}A^a_{\mu}\ =\ 0$, for the $A^a_{\mu}$'s in
(3), then there will be no Gribov ambiguity [22] in fixing the
gauge, in view of (1), as long as we are within the
submanifold, since the gauge variation of the gauge fixing
condition ${\partial}_{\mu}\delta A^a_{\mu}\ =\
{\partial}_{\mu}\ (D^{ab}_{\mu}{\omega}^b)$ identically
vanishes. In this way, the non-propoagating ghost requirement
of t'Hooft [3] is satisfied by the choice of (1). 

\vspace{0.5cm}

{\noindent{\bf{III.SOME PROPERTIES OF THE SUBMANIFOLD.}}}

\vspace{0.5cm}

In this section, we consider the finite gauge transformations
in the submanifold defined by (2),[23]. 
The finite gauge transformations here, will be shown  
become {\it{linear}} in ${\omega}^a$'s. The second relation in
(2) upon using the first relation becomes
$d_{abc}{\omega}^a{\omega}^b{\omega}^c\ =\
\frac{1}{\sqrt{3}}$. It is known [21,24] that, given one octet
vector ${\omega}^a$, there exists at most two linearly
independent octets. They are ${\omega}^a$ itself and
$d^{abc}{\omega}^b{\omega}^c$. Also, it is equally known that
at most two independent $SU(3)$ invariants can be formed. They
are   
${\omega}^a{\omega}^a$ and
$d_{abc}{\omega}^a{\omega}^b{\omega}^c$. In view of (2), we
{\it{now}} have only one octet since
$d^{abc}{\omega}^b{\omega}^c$ is taken to be
$\frac{1}{\sqrt{3}}{\omega}^a$ and the two said $SU(3)$
invariants become constants $1$ and $\frac{1}{\sqrt{3}}$
respectively. These characterize the submanifold.   

\vspace{0.5cm}

Consider a gauge transformation generated by
\begin{eqnarray}
U &=& exp\left(i{\omega}^a \frac{{\lambda}_a}{2}\right),
\end{eqnarray}
for ${\omega}$'s satisfying (2) and ${\lambda}_a$'s are the
Gell-Mann $SU(3)$ matrices. We shall make use of the basic
relation for the $\lambda $- matrices, viz., 
\begin{eqnarray}
{\lambda}_a {\lambda}_b &=& if^{abc}{\lambda}_c +
\frac{2}{3}{\delta}^{ab} + d^{abc}{\lambda}_c,
\end{eqnarray}
where $f^{abc}$ and $d^{abc}$ are the usual anti-symmetric and
symmetric $SU(3)$ tensors respectively. Expanding the
exponential in (7), we have
\begin{eqnarray}
U&=& 1 + i{\omega}^a\frac{{\lambda}_a}{2} - \frac{1}{2!}\
\frac{1}{4}\ ({\omega}^a{\lambda}_a)^2 \nonumber \\
&-& \frac{i}{3!}\ \frac{1}{8}\ ({\omega}^a{\lambda}_a)^3
+ \frac{1}{4!}\ \frac{1}{16}\ ({\omega}^a{\lambda}_a)^4 +
\cdots . \nonumber  
\end{eqnarray}
Using (8), it follows
\begin{eqnarray}
({\omega}^a{\lambda}_a)^2 &=& \frac{2}{3}\ +\
\frac{1}{\sqrt{3}}\ {\omega}^a{\lambda}_a, \nonumber \\
({\omega}^a{\lambda}_a)^3 &=& \frac{2}{3\sqrt{3}} + {\omega}^a
{\lambda}_a, \nonumber \\
({\omega}^a{\lambda}_a)^4 &=& \frac{2}{3} \ +\ \frac{5}{3
\sqrt{3}}\ {\omega}^a{\lambda}_a, \nonumber \\
({\omega}^a{\lambda}_a)^5 &=& \frac{10}{9\sqrt{3}} \ +\ \frac
{11}{9}\ {\omega}^a{\lambda}_a, \nonumber 
\end{eqnarray}
and so on. Substituting these in $U$, we have,
\begin{eqnarray}
U&=& (1-\frac{1}{2!}\ \frac{1}{6}-\frac{i}{3!}\
\frac{1}{12\sqrt{3}}+\frac{1}{4!}\
\frac{5}{144\sqrt{3}}+\cdots ) \nonumber \\
&+&(\frac{i}{2}-\frac{1}{2!}\ \frac{1}{4\sqrt{3}}-\frac{i}
{3!}\ \frac{1}{8}+\frac{1}{4!}\ \frac{5}{48\sqrt{3}}+\cdots )
\ {\omega}^a{\lambda}_a \nonumber \\
&=& \alpha \ +\ \beta \ {\omega}^a{\lambda}_a,
\end{eqnarray}
where $\alpha$ and $\beta$ are (finite) complex constants.
$U^{-1}$ can be easily found to be ${\alpha}^*\ +\ {\beta}^*
{\omega}^a{\lambda}_a$ (where $^*$ stands for complex
conjugation) and so $U{U}^{-1}\ =\ I$, gives the conditions,
\begin{eqnarray}
\alpha {\alpha}^*\ +\ \frac{2}{3}\ \beta {\beta}^* &=& 1,
\nonumber \\
\alpha\ {\beta}^*\ +\ \beta\ {\alpha}^*\ +\ \frac{1}{\sqrt{3}}
\ \beta\ {\beta}^* &=& 0.
\end{eqnarray}
Thus, the finite gauge transformation $U$ becomes
{\it{linear}} in ${\omega}$. This is a special feature of the
submanifold defined by (2). {\it{Thus a parameterization of
$SU(3)$ in terms of a single real octet vector ${\omega}^a$ is
achieved by (9).}}  
It has been shown by Macfarlane,
Sudbery and Weisz [21] that {\it{any}} special unitary matrix $U$
written as $exp(iA)$ with $A$ hermitian and $A\ =\
{\omega}_a{\lambda}_a$, can be written as $U\ =\ u_0 \ +\
iu_a{\lambda}_a$, where $u_a\ =\ a{\omega}_a + b
d_{abc}{\omega}^b{\omega}^c$, with $u_0,a,b$ as functions of
the two $SU(3)$ invariants, ${\omega}^a{\omega}^a$ and
$d_{abc}{\omega}^a{\omega}^b{\omega}^c$. In our choice, these
invariants are taken as constants $1$ and $\frac{1}{\sqrt{3}}$
and this gives (9) for $U$.  
We will relate this $U$ to the
rotation matrix of a subgroup of rotations in eight
dimensional Euclidean space $E8$.   

\vspace{0.5cm}

Now, consider a finite gauge transformation due to $U$ in (9).
Under a finite gauge transformation, we know that the gauge field
$A_{\mu}\ =\ A^a_{\mu}\frac{{\lambda}_a}{2}$ transforms as
\begin{eqnarray}
A_{\mu}\ \rightarrow \ A^U_{\mu}&=& \frac{1}{i}\
({\partial}_{\mu}U)U^{-1} + U\ A_{\mu}\ U^{-1}.
\end{eqnarray}
To check the notation and the factors, we find the
infinitesimal version of (11) using (7), is
\begin{eqnarray}
{\delta}A^a_{\mu} &=& D^{ac}_{\mu}{\omega}^c,
\end{eqnarray}
as it should be. In view (1), $\delta A^a_{\mu}\ =\ 0$. Here we
want to show that for {\it{finite}} gauge transformations (9)
with (10), also $\delta A_{\mu}\ =\ 0$, i.e., 
\begin{eqnarray}
A^{U}_{\mu} &=& A_{\mu},
\end{eqnarray}
for those $A^a_{\mu}$'s satisfying (1). 

\vspace{0.5cm}

We shall make use of the standard relations among $f$ and $d$
tensors [21], viz.,
\begin{eqnarray}
f^{i\ell m}d^{mjk} + f^{j\ell m}d^{imk} + f^{klm}d^{ijm} &=&
0,
\end{eqnarray}
and 
\begin{eqnarray}
f^{ijm}f^{k\ell m}&=&\frac{2}{3}({\delta}^{ik}{\delta}^{j\ell}
- {\delta}^{i\ell}{\delta}^{jk})+d^{ikm}d^{j\ell m}-d^{jkm}
d^{i\ell m}.
\end{eqnarray}

\vspace{0.5cm}

Starting from (9), we have
\begin{eqnarray}
({\partial}_{\mu}U) U^{-1}&=&\beta
({\partial}_{\mu}{\omega}^a){\lambda}_a({\alpha}^*+{\beta}^*
{\omega}^b{\lambda}_b) \nonumber \\
&=&\beta {\alpha}^*({\partial}_{\mu}{\omega}^a){\lambda}_a+
i\beta{\beta}^*({\partial}_{\mu}{\omega}^a){\omega}^bf^{abc}
{\lambda}_c \nonumber \\  
&+&\beta{\beta}^*({\partial}_{\mu}{\omega}^a)
{\omega}^bd^{abc}{\lambda}_c,
\end{eqnarray}
where $({\partial}_{\mu}{\omega}^a){\omega}^a\ =\ 0$,
following from the first relation in (2) is used. Also, the
second relation in (2) gives
\begin{eqnarray}
d^{abc}({\partial}_{\mu}{\omega}^b){\omega}^c &=&
\frac{1}{2\sqrt{3}}\ ({\partial}_{\mu}{\omega}^a),
\end{eqnarray}
and so
\begin{eqnarray}
({\partial}_{\mu}U)U^{-1}&=&
(\beta{\alpha}^*+\frac{1}{2\sqrt{3}}\beta{\beta}^*)
({\partial}_{\mu}
{\omega}^a){\lambda}_a+i\beta{\beta}^*({\partial}_{\mu}
{\omega}^a){\omega}^bf^{abc}{\lambda}_c. 
\end{eqnarray}
We are interested in the transformation of the $A^a_{\mu}$'s
satisfying (1) and so using (1) in (18), we have
\begin{eqnarray}
({\partial}_{\mu}U)U^{-1}&=&(\beta{\alpha}^*+\frac{1}
{2\sqrt{3}}
\beta{\beta}^*)f^{abc}{\omega}^bA^c_{\mu}{\lambda}_a
\nonumber \\
&+& i\beta{\beta}^* (f^{apq}{\omega}^pA^q_{\mu}){\omega}^b
f^{abc}{\lambda}_c.
\end{eqnarray}
Using (15) for simplifying $f^{apq}f^{abc}$, (19) becomes
\begin{eqnarray}
({\partial}_{\mu}U)U^{-1}&=&(\beta{\alpha}^*+\frac{1}{2\sqrt
{3}}\beta{\beta}^*)f^{abc}{\omega}^bA^c_{\mu}{\lambda}_a
\nonumber \\
&+& i\beta{\beta}^*\{ \frac{2}{3}A^a_{\mu}{\lambda}_a -
\frac{2}{3}{\omega}^b{\omega}^cA^b_{\mu}{\lambda}_c + \frac{1}
{\sqrt{3}}{\omega}^a\ d^{qca}A^q_{\mu}{\lambda}_c \nonumber
\\
&-& d^{qba}d^{pca}{\omega}^p{\omega}^bA^q_{\mu}{\lambda}_c\},
\end{eqnarray}
where the relations in (2) have been used.      

\vspace{0.5cm}

Now, we consider the evaluation of $UA_{\mu}U^{-1}$ using (9).
\begin{eqnarray}
UA_{\mu}U^{-1}&=& (\alpha + \beta {\omega}^a{\lambda}_a)\
A^b_{\mu}\frac{{\lambda}_b}{2}\ ({\alpha}^* + {\beta}^*
{\omega}^c{\lambda}_c), \nonumber \\
&=& \frac{1}{2}(\alpha + \beta {\omega}^a{\lambda}_a)\
\{{\alpha}^* {\lambda}_b + i{\beta}^* f^{bcd}\ {\omega}^c 
{\lambda}_d + \frac{2}{3}{\beta}^* {\omega}^b \nonumber \\
&+& {\beta}^* d^{bcd}\ {\omega}^c {\lambda}_d\} A^b_{\mu},
\end{eqnarray}
using (8). Expanding further and using (8), we have
\begin{eqnarray}
UA_{\mu}U^{-1}&=& \frac{1}{2}\{\alpha {\alpha}^*{\lambda}_b +
i\alpha {\beta}^* f^{bcd} {\omega}^c{\lambda}_d + \frac{2}{3}
\alpha {\beta}^* {\omega}^b \nonumber \\
&+&\alpha{\beta}^*d^{bcd}{\omega}^c{\lambda}_d+\beta{\alpha}^*
{\omega}^a(if^{abc}{\lambda}_c+\frac{2}{3}{\delta}_{ab}+
d^{abc}{\lambda}_c) \nonumber \\
&+& i\beta{\beta}^*f^{bcd}{\omega}^c{\omega}^a(if^{ade}
{\lambda}_e+\frac{2}{3}{\delta}_{ad}+d^{ade}{\lambda}_e)
\nonumber \\
&+& \frac{2}{3}\beta{\beta}^*\ {\omega}^a{\omega}^b{\lambda}_a
\nonumber \\
&+&\beta{\beta}^*d^{bcd}{\omega}^a{\omega}^c(if^{ade}{\lambda}
_e+\frac{2}{3}{\delta}_{ad}+d^{ade}{\lambda}_e)\}A^b_{\mu}.
\end{eqnarray} 
In here there are fourteen terms. The sixth term and the third
term are like-terms; the seventh and the fourth are
like-terms, and the fifth and the second are like-terms. The
nineth term vanishes identically. In the thirteenth term, the
second relation in (2) is used. Then we find,
\begin{eqnarray}
UA_{\mu}U^{-1}&=&\frac{1}{2}\{\alpha{\alpha}^*{\lambda}_b+i(
\alpha{\beta}^*-\beta{\alpha}^*)f^{bcd}{\omega}^c{\lambda}_d+
\frac{2}{3}(\alpha{\beta}^*+\beta{\alpha}^*){\omega}^b
\nonumber \\
&+& (\alpha{\beta}^* + \beta{\alpha}^*)
d^{bcd}{\omega}^c{\lambda}_d \nonumber \\
&+&i\beta{\beta}^*(if^{ade}f^{bcd}{\omega}^a{\omega}^c
{\lambda}_e+f^{bcd}d^{ade}{\omega}^c{\omega}^a{\lambda}_e)
\nonumber \\
&+& \frac{2}{3}\beta{\beta}^*{\omega}^a{\omega}^b{\lambda}_a
\nonumber \\
&+&\beta{\beta}^*(if^{ade}d^{bcd}{\omega}^a{\omega}^c{\lambda}
_e+\frac{2}{3\sqrt{3}}{\omega}^b+d^{bcd}d^{ade}{\omega}^a
{\omega}^c{\lambda}_e)\}A^b_{\mu}.
\end{eqnarray} 
Now, the coefficient of ${\omega}^b$ terms in (23) is
$\frac{2}{3}\ (\alpha{\beta}^*\ +\ \beta{\alpha}^*\ +\
\frac{1}{\sqrt{3}}\beta{\beta}^*)$, which is zero due to the
second relation in (10). Using the relation (14) to rewrite the
eighth term $f^{ade}d^{bcd}$ as $-f^{bad}d^{edc}\ -\
f^{cad}d^{ebd}$, the term involving $f^{cad}$ vanishes due to
the symmetry of ${\omega}^a{\omega}^c$ in $a$ and $c$. Then
the remaining of the eighth term cancells with the sixth term
in (23). In the fifth term involving $f^{ade}f^{bcd}$, we make
use of (15). Then (23) becomes,
\begin{eqnarray}
UA_{\mu}U^{-1}&=&\frac{1}{2}\{\alpha{\alpha}^*{\lambda}_b
+i(\alpha{\beta}^*-\beta{\alpha}^*)f^{bcd}{\omega}^c
{\lambda}_d+(\alpha{\beta}^*+\beta{\alpha}^*)d^{bcd}
{\omega}^c{\lambda}_d \nonumber \\
&-&\beta{\beta}^*(\frac{2}{3}{\lambda}_b-\frac{2}{3}{\omega}
^b{\omega}^c{\lambda}_c+\frac{d^{ebd}}{\sqrt{3}}{\omega}^d
{\lambda}_e-d^{abd}d^{ecd}{\omega}^a{\omega}^c{\lambda}_e)
\nonumber \\
&+&\frac{2}{3}\beta{\beta}^*{\omega}^a{\omega}^b{\lambda}_a
+\beta{\beta}^*d^{bcd}d^{ade}{\omega}^a{\omega}^c{\lambda}
_e\}A^b_{\mu}.
\end{eqnarray} 
Here, the two terms involving the two $d$-tensors are the same
and so,
\begin{eqnarray}
UA_{\mu}U^{-1}&=&\frac{1}{2}\{\alpha{\alpha}^*{\lambda}_b
+i(\alpha{\beta}^*-\beta{\alpha}^*)f^{bcd}{\omega}^c
{\lambda}_d+(\alpha{\beta}^*+\beta{\alpha}^*)d^{bcd}
{\omega}^c{\lambda}_d \nonumber \\
&-&\frac{2}{3}\beta{\beta}^*{\lambda}_b+\frac{4}{3}\beta
{\beta}^*{\omega}^a{\omega}^b{\lambda}_a-\beta{\beta}^*
\frac{1}{\sqrt{3}}d^{ebd}{\omega}^d{\lambda}_e \nonumber \\
&+& 2\beta{\beta}^* d^{abd}d^{ecd}{\omega}^a{\omega}^c
{\lambda}_e\}\ A^b_{\mu}.
\end{eqnarray}
From (20) and (25), we have,
\begin{eqnarray}
\frac{1}{i}({\partial}_{\mu}U)U^{-1}+UA_{\mu}U^{-1}&=&-i(\beta
{\alpha}^*+\frac{1}{2\sqrt{3}}\beta{\beta}^*)f^{abc}{\omega}
^b{\lambda}_aA^c_{\mu} \nonumber \\
&+&\beta{\beta}^*\{\frac{2}{3}A^a_{\mu}{\lambda}_a-\frac{2}
{3}{\omega}^b{\omega}^cA^b_{\mu}{\lambda}_c \nonumber \\
&+&\frac{1}{\sqrt{3}}d^{qca}{\omega}^aA^q_{\mu}{\lambda}_c
-d^{qba}d^{pca}{\omega}^p{\omega}^bA^q_{\mu}{\lambda}_c\}
\nonumber \\
&+&\frac{1}{2}\{\alpha{\alpha}^*{\lambda}_b+i(\alpha{\beta}^*
-\beta{\alpha}^*)f^{bcd}{\omega}^c{\lambda}_d \nonumber \\
&+&(\alpha{\beta}^*+\beta{\alpha}^*)d^{bcd}{\omega}^c
{\lambda}_d-\frac{2}{3}\beta{\beta}^*{\lambda}_b+\frac{4}{3}
\beta{\beta}^*{\omega}^a{\omega}^b{\lambda}_a \nonumber \\
&-&\frac{\beta{\beta}^*}{\sqrt{3}}d^{ebd}{\omega}^d{\lambda}
_e+2\beta{\beta}^*d^{abd}d^{ecd}{\omega}^a{\omega}^c{\lambda}
_e\}A^b_{\mu}.
\end{eqnarray}  
 
\vspace{0.5cm}

Now, consider the terms involving two $d$-tensors. They are,
\begin{eqnarray}
(-d^{qba}d^{pca}{\omega}^p{\omega}^bA^q_{\mu}{\lambda}_c\ +\
d^{abd}d^{ecd}{\omega}^a{\omega}^cA^b_{\mu}{\lambda}_c).
\nonumber 
\end{eqnarray}
By changing the summation indices in the second term above,
viz., $b\rightarrow q, \ a\rightarrow b,\ c\rightarrow p,\
e\rightarrow c$ in that order, the two terms cancell each
other. Then, the terms involving one $d$-tensor are,
\begin{eqnarray}
\beta{\beta}^*d^{qca}{\omega}^aA^q_{\mu}{\lambda}_c+\frac{1}
{2}(\alpha{\beta}^*+\beta{\alpha}^*)d^{bcd}{\omega}^c
{\lambda}_dA^b_{\mu}-\frac{\beta{\beta}^*}{2\sqrt{3}}d^{ebd}
{\omega}^d{\lambda}_eA^b_{\mu}. \nonumber  
\end{eqnarray}
By rearranging the indices, this becomes,
\begin{eqnarray}
\frac{1}{2}(\alpha{\beta}^*+\beta{\alpha}^*+\frac{1}{\sqrt{3}}
\beta{\beta}^*)d^{qca}{\omega}^aA^q_{\mu}{\lambda}_c,
\nonumber 
\end{eqnarray}
which vanishes due to the second relation in (10). Now,
consider the terms involving one $f$-tensor. They are,
\begin{eqnarray}
-i(\beta{\alpha}^*+\frac{\beta{\beta}^*}{2\sqrt{3}})f^{abc}
{\omega}^bA^c_{\mu}{\lambda}_a+\frac{i}{2}(\alpha{\beta}^*
-\beta{\alpha}^*)f^{bcd}{\omega}^c{\lambda}_dA^b_{\mu}.
\nonumber 
\end{eqnarray}
In the second term, make $b\leftrightarrow c,\ d\rightarrow a$
so that the above expression becomes,
\begin{eqnarray}
-\frac{i}{2}(\alpha{\beta}^*+\beta{\alpha}^*+\frac{1}{\sqrt{3}
}\beta{\beta}^*)f^{abc}{\omega}^bA^c_{\mu}{\lambda}_a,
\nonumber 
\end{eqnarray}
which vanishes due to the second relation in (10). Finally, the
terms involving two ${\omega}$'s in (26) cancel each other,
leaving
\begin{eqnarray}
\frac{1}{i}({\partial}_{\mu}U)U^{-1}+UA_{\mu}U^{-1}&=&
\frac{1}{2} (\alpha{\alpha}^*+\frac{2}{3}\beta{\beta}^*)
A^a_{\mu}{\lambda}_a\ =\ \frac{1}{2}A^a_{\mu}{\lambda}_a,
\nonumber \\
&=& A_{\mu},
\end{eqnarray}
using the first relation in (10).

\vspace{0.5cm}

Thus when $U\ =\ exp\{i{\omega}^a\frac{{\lambda}_a}{2}\}$
{\it{finite}} transformation is considered, $U$ becomes
$\alpha +\beta {\omega}^a{\lambda}_a$,  {\it{linear}} in
${\omega}^a$ in view of (2), with $\alpha$ and $\beta$
satisfying (10). This finite transformation leaves $A_{\mu}$
invariant, for those $A^a_{\mu}$'s satisfying (1). Therefore,
within the submanifold defined by (2), the gauge fields
satisfying (1) can have a mass term, even for finite
transformations.

\vspace{0.5cm}

It will be useful to consider the converse of this result. In
order to this, we ask the question: What will be the gauge
field configurations that remain unchanged under finite gauge
transformations? Since under a finite gauge transformation
$U$, $A_{\mu}$ transforms as,
\begin{eqnarray}
A_{\mu}\ \rightarrow \ A^U_{\mu} &=&
\frac{1}{i}({\partial}_{\mu}U)U^{-1} + UA_{\mu}U^{-1},
\nonumber
\end{eqnarray}
the answer is given by
\begin{eqnarray}
A_{\mu} &=&
\frac{1}{i}({\partial}_{\mu}U)U^{-1}+UA_{\mu}U^{-1}.  \nonumber
\end{eqnarray}
By right multiplying the above expression by $U$, we have
\begin{eqnarray}
\frac{1}{i}{\partial}_{\mu}U + [U,A_{\mu}] &=& 0,
\end{eqnarray}
whose infinitesimal version is (1). This means that for those
$A_{\mu}$'s and $U$ satisfying (28), the mass term remain
invariant. In order to give explicit expressions for
$A^a_{\mu}$ and $U$, the transformation $U$ is chosen by those
${\omega}$'s satisfying (2). Then (9) gives the required $U$.
The choice of ${\omega}$'s satisfying (2) and $A^a_{\mu}$ in
(3) determined by (1),  
produce magnetic
monopole configurations in the QCD action [12].  

\vspace{0.5cm}

It is known that the tensor indices taking eight values in $d$
and $f$ are tensor indices associated with the adjoint group
$SU(3)/Z(3)$ of $SU(3)$. The tensors $d$ and $f$ associated
with $SU(3)/Z(3)$, are cartesian tensors in eight real
dimensions [24]. Given a single octet vector $\{ {\omega}^a
\}$ and the tensors $d$ and $f$, it is known that {\it{at
most}} two linearly independent octets can be formed. They are
${\omega}^a$ itself and $d_{abc}{\omega}^b{\omega}^c$. Then
{\it{at most}} two independent $SU(3)$ invariants can be
formed, which are taken as, 
\begin{eqnarray}
{\omega}^a{\omega}^a, \nonumber \\
d_{abc}{\omega}^a{\omega}^b{\omega}^c.
\end{eqnarray} 
A geometric meaning can be given to the invariants. If we
associate a $3\times 3$ matrix $A$ with ${\omega}^a$ as $A\
=\ {\omega}^a\ {\lambda}_a$, where ${\lambda}^a$'s are the
Gell-Mann matrices, then $Tr(A^2)$ and $det(A)$ give the above
two invariants. In our choice made in (2), we have taken the
two invariants as constants. 

\vspace{0.5cm}

A study of the relationship of the adjoint group $SU(3)/Z(3)$
to the subgroup of rotation group $R8$ which leaves invariant
the length ${\omega}^a{\omega}^a$ of the real eight component
vector ${\omega}^a$ and the cubic invariant
$d_{abc}{\omega}^a{\omega}^b{\omega}^c$ has been made by
Macfarlane [24]. Here we will give the main results for our
choice (2). The eight components ${\omega}^a$ can be taken to
describe a point of $E8$. Rotations in $E8$ are real linear
transformations
\begin{eqnarray}
{\omega}^a\ \rightarrow \ {\omega}^{'a} &=& R_{ab}{\omega}^b.
\end{eqnarray}
Invariance of the length leads to $RR^T\ =\ I$. We take
$det(R)\ =\ 1$. To relate $SU(3)$ to a subgroup of $R8$, the
group of rotations in $E8$, associate with each point in $E8$,
a $3\times 3$ traceless hermitian matrix A
\begin{eqnarray}
A &=& {\omega}^a{\lambda}_a,
\end{eqnarray}
an element of the algebra of $SU(3)$. Transformations of $E8$
induced by $U\ \in \ SU(3)$ tansformation
\begin{eqnarray}
A\ \rightarrow\ A' &=& UAU^{-1},
\end{eqnarray}
can be shown to give [24] 
\begin{eqnarray}
R_{ab} &=& \frac{1}{2}\ Tr\{ {\lambda}_a \ U\ {\lambda}_b\
U^{-1}\}.
\end{eqnarray}
For $U$ in (9) and using the relations in (10), we find
explicitly,
\begin{eqnarray}
R_{ab}&=&(1-{\beta}{\beta}^*){\delta}_{ab}+2{\beta}{\beta}^*
{\omega}_a{\omega}_b-\sqrt{3}{\beta}{\beta}^*d_{abc}{\omega}
_c \nonumber \\
&+& if_{abc}{\omega}_c\ (\alpha{\beta}^*-\beta{\alpha}^*).
\end{eqnarray}
It is verified that $R_{ab}R_{ac}\ =\ {\delta}_{bc}$, i.e.,
the transformation is orthogonal. It is seen that
$R_{ab}{\omega}_b\ =\ {\omega}_a$. Thus the rotation leaves
${\omega}^a{\omega}^a$ invariant. It is seen that the
invariance of $d_{abc}{\omega}^a{\omega}^b{\omega}^c$ is
verified by showing $d_{abc}R_{ap}R_{bq}R_{cr}\ =\ d_{pqr}$.

\vspace{0.5cm}

Now, we examine the property of an octet $B\ =\
{\omega}^a{\lambda}_a$ in the submanifold. This will shed some
light on the type of monopole configurations discussed in the
next section. In order for this, we consider the
"characteristic equation" for the $3\times 3$ matrix $B$. It
is [21] (for the choices in (2), namely the two $SU(3)$
invariants taken as $1$ and $\frac{1}{\sqrt{3}}$, constants.) 
given by $x^3\ -\ x\ -\ \frac{2}{3\sqrt{3}}\ =\ 0$
with $x$ as the eigenvalues of $B$. The eigenvalues are found
to be $-\frac{1}{\sqrt{3}},\ -\frac{1}{\sqrt{3}},\
\frac{2}{\sqrt{3}}$, {\it{showing that in its diagonalized
form, $B$ is $-{\lambda}_8$-like.}} In this way, the monopole
configurations in section.IV, will be ${\lambda}_8$-like and
the submanifold defined by (2), picks up this configuration.
It is possible to realize the ${\lambda}_3$-like
configuration, by going out of the submanifold, which we will
not do in this study.  

\vspace{0.5cm}

{\noindent{\bf{IV.LOW ENERGY EFFECTIVE ACTION.}}}

\vspace{0.5cm}

The field strength $F^a_{\mu\nu}\ =\
{\partial}_{\mu}A^a_{\nu}-{\partial}_{\nu}A^a_{\mu}+g f^{abc}
A^b_{\mu}A^c_{\nu}$ for the $A^a_{\mu}$ in (3) is calculated
as
\begin{eqnarray}
F^a_{\mu\nu}&=&
({\partial}_{\mu}C_{\nu}-{\partial}_{\nu}C_{\mu})\ {\omega}^a
- \frac{8}{3g}f^{abc}({\partial}_{\mu}{\omega}^b)({\partial}_
{\nu}{\omega}^c) \nonumber \\
&+& \frac{16}{9g}f^{abc}f^{bed}f^{c\ell m}\ {\omega}^e
{\omega}^{\ell}\ {\partial}_{\mu}{\omega}^d\ {\partial}_{
\nu}{\omega}^m.
\end{eqnarray} 
Use of the Jacobi identity for $f$'s [21] and the relation (6)
for the last term in (35) gives
\begin{eqnarray}
F^a_{\mu\nu}&=&({\partial}_{\mu}C_{\nu}-{\partial}_{\nu}C_
{\mu})\ {\omega}^a - \frac{4}{3g}f^{abc}\ ({\partial}_{\mu}
{\omega}^b) ({\partial}_{\nu}{\omega}^c) \nonumber \\
&-& \frac{16}{9g}f^{a\ell c}f^{bed}f^{mbc}\ {\omega}^e
\ {\omega}^{\ell} ({\partial}_{\mu}{\omega}^d) ({\partial}
_{\nu}{\omega}^m).
\end{eqnarray}
Now using $f^{edb}f^{cmb}\ =\ \frac{2}{3}({\delta}_{ec}
{\delta}_{dm}-{\delta}_{em}{\delta}_{ed})+d^{ecb}d^{dmb}
-d^{dcb}d^{emb}$ and the relations (2), the last term in (36)
can be shown to vanish, leaving
\begin{eqnarray}
F^a_{\mu\nu}&=& ({\partial}_{\mu}C_{\nu}-{\partial}_{\nu}C_
{\mu})\ {\omega}^a - \frac{4}{3g}f^{abc}\ ({\partial}_{\mu}
{\omega}^b) ({\partial}_{\nu}{\omega}^c). 
\end{eqnarray}
Consistent with (1), the above field strength $F^a_{\mu\nu}$
is ``$SU(3)$ parallel" to ${\omega}^a$ i.e.,
\begin{eqnarray}
f^{abc}\ {\omega}^b\ F^c_{\mu\nu}&=& 0,
\end{eqnarray}
which can be verified by using (4) and the relation among
$f$'s. Unlike the case of $SU(2)$ [10,11], this does not imply
that $F^a_{\mu\nu}$ is along ${\omega}^a$.

\vspace{0.5cm}

The QCD action in the absense of quarks,
\begin{eqnarray}
S &=& -\frac{1}{4}\int d^4x (F^a_{\mu\nu})^2,
\end{eqnarray}
becomes
\begin{eqnarray}
S &=& -\frac{1}{4}\int d^4x \{ f^2_{\mu\nu} - \frac{8}{3g}
f_{\mu\nu}\ (f^{abc}{\omega}^a\ {\partial}_{\mu}{\omega}^b\ 
 {\partial}_{\nu}{\omega}^c) \nonumber \\
&+& \frac{16}{9g^2}f^{abc}f^{aed}\ {\partial}_{\mu}{\omega}^b
\ {\partial}_{\nu}{\omega}^c\ {\partial}_{\mu}{\omega}^e\ 
 {\partial}_{\nu}{\omega}^d\},
\end{eqnarray}
when (37) is used and where 
\begin{eqnarray}
f_{\mu\nu}& =& {\partial}_{\mu}
C_{\nu} - {\partial}_{\nu}C_{\mu}.
\end{eqnarray}   
It will be convenient to rescale the ${\omega}^a$'s as
$g^{\frac{1}{3}}\ {\omega}^a$ and then in the strong coupling
limit, the above action becomes
\begin{eqnarray}
S &\simeq & -\frac{1}{4}\int d^4x \{ f^2_{\mu\nu} -
\frac{8}{3}f_{\mu\nu}\ (f^{abc}{\omega}^a\ {\partial}_{\mu}
{\omega}^b\ {\partial}_{\nu}{\omega}^c)\},
\end{eqnarray}
showing the Abelian dominance in the infrared region of QCD.
Denoting $f^{abc}\ {\omega}^a\ {\partial}_{\mu}{\omega}^b\ 
{\partial}_{\nu}{\omega}^c\ =\ X_{\mu\nu}$, we see that
$f_{\mu\nu}$ is coupled to $X_{\mu\nu}$. It follows that
${\partial}_{\mu}X_{\mu\nu}\ =\ f^{abc}{\omega}^a\ 
{\partial}_{\mu}({\partial}_{\mu}{\omega}^b\ {\partial}_{\nu}
{\omega}^c)\ \neq \ 0$, and the dual $\bar{X}_{\mu\nu}\ =\
\frac{1}{2}{\epsilon}_{\mu\nu\alpha\beta}X_{\alpha\beta}$
violates the Bianchi idendity,
\begin{eqnarray}
{\partial}_{\mu}\bar{X}_{\mu\nu}&=& \frac{1}{2}{\epsilon}_
{\mu\nu\alpha\beta}\ f^{abc}\ {\partial}_{\mu}{\omega}^a
{\partial}_{\alpha}{\omega}^b {\partial}_{\beta}{\omega}^c
\ \neq \ 0.
\end{eqnarray}
This along with
\begin{eqnarray}
-\frac{2}{3}\oint_{s} {\epsilon}_{\mu\nu\alpha\beta}\ f^{abc}
{\omega}^a\ {\partial}_{\alpha}{\omega}^b\ {\partial}_{\beta}
{\omega}^c\ dx^{\mu}\wedge dx^{\nu}&=& -\frac{4}{3}\oint_{s}
\bar{X}_{\mu\nu}\ dx^{\mu}\wedge dx^{\nu},
\end{eqnarray}
a topological invariant, imply that magnetic monopoles are
present in (42). In this way, the proposed gauge field
configurations (3) give an action (42) which contains the
monopole configurations. In the attempts [9] to relate the
infrared region of QCD to a non-linear sigma model by
expressing $A^a_{\mu}$ in terms $n^a$ (the sigma model
fields), it is to be {\it{observed}} that the QCD action (39)
will produce {\it{quartic}} terms involving the derivatives of
$n^a$. The {\it{quadratic}} term in the sigma model action,
namely, ${\partial}_{\mu}n^a{\partial}_{\mu}n^a$ can come only
from the mass term $A^a_{\mu}A^a_{\mu}$. So, such models [9]
implicitly assume the presence of a mass term for $A^a_{\mu}$
fields in the QCD lagrangian. Such a mass term, if included,
will induce a mass term for the $C_{\mu}$ fields in our
approach [12].  
In the partition function $Z$ for the
action (42), after introducing a mass term for the
$C_{\mu}$-field with mass $m$  
(for instance by Coleman-Weinberg mechanism by
introducing complex scalars minimally coupled to $C_{\mu}$ as
in [5,7,11]), a functional integration over $C_{\mu}$ produces
an effective lagrangian [12]
\begin{eqnarray}
{\cal{L}}_{eff} &=& -\frac{8}{9}\ {\partial}_{\mu}X_{\mu\nu}\
(\Box - \frac{m^2}{2})^{-1} {\partial}_{\rho}X_{\rho\nu}, 
\end{eqnarray}
a form identical to the London case of magnetic confinement,
as in an ordinary superconductor due to Meissner effect.    

\vspace{0.5cm}

{\noindent{\bf{V. A STRING REPRESENTATION OF THE EFFECTIVE
ACTION.}}}

\vspace{0.5cm}

In order to find a string representation of the field
theoretic action (42), we first consider its dual form. Introducing
the dual field strength ${\cal{G}}_{\mu\nu}$ dual to
$f_{\mu\nu}$, we have
\begin{eqnarray}
Z &=& \int [dC_{\mu}][{\cal{G}}_{\mu\nu}] exp\{\int
(-\frac{1}{4}{\cal{G}}^2_{\mu\nu}+\frac{1}{2}{\cal{G}}_{
\mu\nu}f_{\mu\nu}-\frac{2}{3}f_{\mu\nu}X_{\mu\nu})d^4x\}.
\end{eqnarray}
The functional integral over ${\cal{G}}_{\mu\nu}$ produces the
partition function for (42). From (46), variation with respect
to the $C_{\mu}$-field ($f_{\mu\nu}\ =\
{\partial}_{\mu}C_{\nu}-{\partial}_{\nu}C_{\mu}$) gives
${\partial}_{\mu}\{{\cal{G}}_{\mu\nu}-\frac{4}{3}X_{\mu\nu}\ 
\} =\ 0$, which is solved for ${\cal{G}}_{\mu\nu}$ as 
\begin{eqnarray}
{\cal{G}}_{\mu\nu} &=& {\epsilon}_{\mu\nu\lambda\sigma}
{\partial}_{\lambda}{\bar{A}}_{\sigma}\ +\ \frac{4}{3}X_
{\mu\nu},
\end{eqnarray}
where the field ${\bar{A}}_{\mu}$ serves as dual to $C_{\mu}$.
Using (47) in (46), we eliminate the $C_{\mu}$-field to obtain
\begin{eqnarray}
Z&=& \int
[d\bar{A}_{\mu}]exp[\frac{1}{4}\{({\partial}_{\lambda}{\bar{A}
}_{\sigma}-{\partial}_{\sigma}{\bar{A}}_{\lambda})^2+\frac{8}
{3}X_{\mu\nu}{\epsilon}_{\mu\nu\lambda\sigma}{\partial}_
{\lambda}{\bar{A}}_{\sigma} \nonumber \\
&+&\frac{16}{9}X_{\mu\nu}X_{\mu\nu}\}d^4x].
\end{eqnarray}
This action posseses dual $U(1)$ invariance and coincides with
the Abelian projected effective theory of QCD based on $SU(3)$
in its dual form. We introduce a mass term for the
$\bar{A}_{
\sigma}$-field which can arise from Coleman-Weinberg mechanism
by invoking complex scalars coupled to $\bar{A}_{\sigma}$ as
in the works of [3,4,11]. By functionally integrating the
$\bar{A}_{\sigma}$-field, we obtain an effective action
\begin{eqnarray}
S_{eff}&=& \int \{-\frac{8}{9}{\partial}_{\lambda}{\bar{X}}_
{\lambda\sigma}(\Box - \frac{m^2}{2})^{-1}\ {\partial}_
{\rho}{\bar{X}}_{\rho\sigma} - \frac{4}{9}{\bar{X}}_
{\mu\nu}{\bar{X}}_{\mu\nu}\}\ d^4x,
\end{eqnarray}
apart from a constant (divergent) factor not involving the
fields. This action suggests a dual confinement of gluons in
the submanifold as in the London theory of Meissner effect.   
Second, we go over now to the Euclidean space. 

\vspace{0.5cm}

From (43) and (44) it follows that $\bar{X}_{\mu\nu}$
represents the field strength of magnetic monopole. We make
the following choice for $X_{\mu\nu}$,
\begin{eqnarray}
X_{\mu\nu}(x)&=& \int d^2\xi {\delta}^4(x-y)[y_{\mu},y_{\nu}],
\end{eqnarray}
where $[y_{\mu},y_{\nu}]\ =\ {\epsilon}^{ab}\
\frac{{\partial}y_{\mu}}{{\partial}{\xi}^a}\ \frac{{\partial}
y_{\nu}}{{\partial}{\xi}^b}$ ; $a,b\ =\ 1,2$,
$y_{\mu}({\xi}_1,{\xi}_2)$ represents the position of a point
on the worldsheet swept by the string of the monopole and
${\xi}_1,\ {\xi}_2$ are the local coordinates of the
worldsheet. In here, we have the Nambu's picture [6] in mind
that the magnetic flux lines terminate at the end points
(quarks). Such a form for $X_{\mu\nu}$ has been earlier
suggested by Wentzel [25]. Quarks exchange the gluons and in
the infrared region, the proposed gauge field configuration
(3) produces an action (42) which contains $X_{\mu\nu}$. Thus
it is consistent with the picture that the magnetic flux lines
connect two quarks or quark - antiquark pair. 

\vspace{0.5cm}

Now using ${\bar{X}}_{\mu\nu}\ =\
\frac{1}{2}{\epsilon}_{\mu\nu\alpha\beta}\ X_{\alpha\beta}$,
it is found that
\begin{eqnarray}
\int {\partial}_{\lambda}{\bar{X}}_{\lambda\sigma}(\Box -
\frac{m^2}{2})^{-1}{\partial}_{\rho}{\bar{X}}_{\rho\sigma}d^4x
&=& -\frac{1}{2}\int
X_{\alpha\beta}X_{\alpha\beta}d^4x \nonumber \\  
&-& \frac{m^2}{4}\int X_
{\alpha\beta}(\Box - \frac{m^2}{2})^{-1}X_{\alpha\beta}d^4x,
\end{eqnarray}
and ${\bar{X}}_{\mu\nu}{\bar{X}}_{\mu\nu}\ =\
X_{\mu\nu}X_{\mu\nu}$. Then Eqn.49 becomes, (Euclidean
version),
\begin{eqnarray}
S_{eff} &=&- \frac{2m^2}{9} \int X_{\mu\nu}(\Box -
\frac{m^2}{2})^{-1}X_{\mu\nu} d^4x.
\end{eqnarray}
It is to be noted that the second term on the right hand side
of Eqn.49 is exactly cancelled by the first term in (51). We
now use the representation for $X_{\mu\nu}$  (50) and denote
$[y_{\mu},y_{\nu}]\ =\ {\sigma}_{\mu\nu}(y)$. It is easy to
verify that
\begin{eqnarray}  
{\sigma}_{\mu\nu}(x){\sigma}_{\mu\nu}(x) &=& 2g,
\end{eqnarray}
where $g$ is the determinant of the induced metric $g_{ab}\
=\ {\partial}_aX^{\mu}{\partial}_bX^{\mu}$ on the string
worldsheet. The operator $({\Box}_x - \frac{m^2}{2})^{-1}$
acting on ${\delta}^4(x-y)$ can be evaluated using the
procedure outlined by Diamantini, Quevedo and Trugenberger
[15], by derivative expansion of the Green function of the
operator above. Introducing an ultra-violet cut-off $\Lambda$
(corresponding to finite thickness of the worldsheet), the
above action (52) becomes 
\begin{eqnarray}
S_s &=& \frac{2}{9}\ \frac{m^2}{4\pi}\
K_0(\frac{m}{2\Lambda})\int \sqrt{g} d^2\xi - \frac{2}{9}\
\frac{{\Lambda}^2}{8\pi m^2}\int \sqrt{g} g^{ab}{\partial}_a
t_{\mu\nu}\ {\partial}_bt_{\mu\nu}\ d^2\xi \nonumber \\
&+&- \frac{2}{9}\ \frac{{\Lambda}^2}{m^2}\ \int \sqrt{g}R
\ d^2\xi,
\end{eqnarray}
where $t_{\mu\nu} \ =\ \frac{1}{\sqrt{g}}{\sigma}_{\mu\nu}$
and $R$ is the scalar curvature of the worldsheet. Here the
first term is identified with the area term or the Nambu-Goto
action. The second term is the familiar extrinsic curvature
action (see the book by A.M.Polyakov in Ref.17, page.284) and
the last term is the topological Euler characteristic of the
worldsheet. Since the action (52) has no coupling parameters in
the strong coupling limit, the coefficients in front of the
second third terms are numbers while the coefficient in front
of the first term represents the string tension. The divergent
integrals in the ultra-violet cut-off momentum $\Lambda$ are
approximated by $\frac{{\Lambda}^2}{m^2}$. Eqn.54 represents
the string action for the infrared region of QCD and as we
have shown a realization of confinement as in the London
theory of Meissner effect, the above action is the confining
string action. This agrees with Ref.15, although they start
from Kalb-Ramond action with monopoles while we obtain (54)
from pure QCD action in the infrared region characterized by
the submanifold (2).  

\vspace{0.5cm}

{\noindent{\bf{VI. Conclusions}}}

\vspace{0.5cm}

$SU(3)$ gauge field configurations appropriate for the
infrared region of QCD are proposed in submanifold of $su(3)$.
The usual action of QCD then contains monopole configurations
interacting with $f_{\mu\nu}$. A dual version is constructed
and is shown to give the confinement of these field
configurations as in Meissner effect. A string representation
is obtained by choosing a representation for $X_{\mu\nu}$ in
terms of the geometry of the string worldsheet and by going
over to the Euclidean version. 

\vspace{0.5cm}

The string
action which corresponds to confinement consists of the
Nambu-Goto term, extrinsic curvature action and the Euler
characteristic of the worldsheet, showing the inevitable
occurrence of the extrinsic curvature action. In this
connection between the gauge field theory and the string
theory,  
the origin of
the extrinsic curvature action is in the derivative expansion
of the Green function for the  four dimensional operator $(\Box -
\frac{m^2}{2})^{-1}$ acting on $X_{\mu\nu}$ which contains the
4-dimensional Dirac delta function. The leading term however
gives the Nambu-Goto action. Since the gauge field action has
been shown to have confinement of these gauge fields, one can
expect that the resulting string action also to exhibit
confinement in the infrared region. This has been shown by an
explicit calculation of the quantum one-loop partition
function of both the Nambu-Goto action and extrinsic curvature
action by Viswanathan and Parthasarathy [26], in which the
partition function has been found to be that of a modified
Cpulomb gas and in the infrared limit, there are long range
interaction responsible for confinement while in the
ultra-violet limit, the flux lines curl up to vortices leading
to the non-confining phase of the system. 

\vspace{0.5cm}

Having realized a mechanism of confinement in the infrared
region of QCD, it is pertinent to address the issue of chiral
symmetry breaking as these two issues are related. It has
already been shown by Nair [27] and Nair and Rosenzweig [28]
that quarks in the monopole background break chiral symmetry,
i.e., a non-vanishing value for $<\bar{\psi}\psi>$ has been
obtained in the monopole background, the mechanism being
similar to that of Callan [29] and Rubakov [30], except that
in QCD, this does not violate flavour number. As we have a
monopole configuration realized in (42), by choosing a
specific form for $X_{\mu\nu}$, chiral symmetry breaking can
be realized along the lines of [27] and [28]. 

\vspace{0.5cm}

{\noindent{\bf{Acknowledgements.}}}

Useful discussions with Ramesh Anishetty, H.Sharatchandra,
G.Rajasekaran and Biswajit Chakraborty are acknowledged with
thanks. 

\vspace{0.5cm}

{\noindent{\bf{References.}}}

\vspace{0.5cm}

\begin{enumerate}

\item H.D.Politzur, Phys.Rev.Lett. {\bf 30} (1973) 1346. \\
      D.J.Gross and F.Wilczek, Phys.Rev.Lett. {\bf 30} (1973)
1343. \\
      G.t'Hooft, (unpublished) 1972; See hep-th/9812203.

\item T.Greenshaw, M1 Colloboration and A.Doyle. ZEUS
colloboration in the Proceedings of the Sixth International
Workshop in DIS and QCD, April 1998; Brussels, Belgium.

\item G.t'Hooft, Nucl.Phys. {\bf B190} (1981) 455. 

\item S.Mandelstam, Phys.Rev. {\bf D19} (1978) 2391. 

\item G.t'Hooft, in {\it High Energy Physics Proceedings}, 1975,
Edited by A.Zichichi; Nucl.Phys. {\bf B138} (1978)1; {\bf
B153} (1979) 141. 

\item Y.Nambu, Phys.Rev. {\bf D10} (1974) 4262; Phys.Rep. {\bf
C23} (1975) 250.

\item S.Mandelstam, in the {\it Proceedings of the Monopole
Meeting}, Trieste, Italy, 1981. Edited by N.S.Craige,
P.Goddard and W.Nahm; World SCientific. 1982. 

\item K.-I.Kondo, Phys.Rev. {\bf D58} (1998) 105016, 105019.

\item L.Faddeev and A.J.Niemi, hep-th/9807069; 9812090.

\item E.Corrigan and D.Olive, Nucl.Phys. {\bf B110} (1976)
237. \\
      E.Corrigan, D.Olive,D.B.Fairlie and J.Nuyts, Nucl.Phys.
{\bf B106} (1976) 475.

\item Y.M.Cho, Phys.Rev. {\bf D21} (1980) 1080; {\bf D23}
(1981) 2415; Phys.Rev.Lett. {\bf 44} (1980) 1115. 

\item R.Parthasarathy, hep-th/9902027.

\item A.M.Polyakov, Nucl.Phys. {\bf B486} (1997) 23. 

\item A.M.Polyakov, Phys.Lett. {\bf B59} (1975) 82. 

\item M.C.Diamantini, F.Quevedo and C.A.Trugenberger,
Phys.Lett. {\bf B396} (1997) 115. 

\item M.Kalb and P.Ramond, Phys.Rev. {\bf D9} (1974) 2273. 

\item A.M.Polyakov, Nucl.Phys. {\bf B268} (1986) 406. \\
      A.M.Polyakov, {\it Gauge Fields and Strings}, Harwood
Academic Publishers, Chur, 1987.

\item H.Kleinert, Phys.Lett. {\bf B174} (1986) 335;
Phys.Rev.Lett. {\bf 58} (1987) 1915. 

\item H.Kleinert, hep-th/9601030.

\item H.Kleinert and A.M.Chervyakov, Phys.Lett. {\bf B381}
(1996) 286. 

\item A.J.Macfarlane, A.Sudbury and P.H.Weise, Comm.Math.Phys.
{\bf 11} (1968) 77. 

\item V.N.Gribov, Nucl.Phys. {\bf B139} (1978) 1. 

\item R.Parthasarathy, hep-th/9903060.

\item A.J.Macfarlane, Comm.Math.Phys. {\bf{11}}, (1968) 91. 

\item G.Wentzel, Supplement of Prog.Theor.Phys. Nos. {\bf
37,38} (1966) 163.

\item K.S.Viswanathan and R.Parthasarathy, Phys.Rev. {\bf D51}
(1995) 5830.

\item V.P.Nair, Phys.Rev. {\bf{D28}}, (1983) 2673. 

\item V.P.Nair and C.Rosenzweig, Phys.Lett. {\bf{131B}}, (1983)
434; Phys.Lett. {\bf{135B}}, (1984) 450; Phys.Rev. {\bf{D31}},
(1985) 401.

\item C.Callan, Phys.Rev. {\bf{D25}}, (1982) 2141; Phys.Rev.
{\bf{D26}}, (1982) 2058.

\item V.Rubakov, JETP Lett. {\bf{33}}, (1981) 644; Nucl.Phys.
{\bf{B203}}, (1982) 2058. 
\end{enumerate}

\end{document}